\documentclass{fic-l}
\usepackage{amssymb,amscd}
\def\copyrightyear{2000}
\def\currentvolume{27}
\def\currentyear{2000}
\newtheorem{theorem}{Theorem}[section]
\newtheorem{lemma}[theorem]{Lemma}
\newtheorem{corollary}[theorem]{Corollary}
\theoremstyle{definition}
\newtheorem{definition}[theorem]{Definition}
\newtheorem{example}[theorem]{Example}
\newtheorem{algorithm}[theorem]{Algorithm}

\theoremstyle{remark}
\numberwithin{equation}{section}

\newcommand{\ZZ}{{\mathbb{Z}}}
\newcommand{\NN}{{\mathbb{N}}}
\newcommand{\CG}{{\mathcal G}}
\newcommand{\CB}{{\mathcal B}}
\newcommand{\no}{\bullet}
\begin{document}
\title{A class of cellular automata equivalent to
          deterministic particle systems}
\author{Henryk Fuk\'s}
\address{
Department of Mathematics\\
 Brock University\\
St. Catharines, Ontario L2S 3A1, Canada\\
{\texttt hfuks@brocku.ca}
         }
\thanks{
The author wishes to thank The Fields Institute for Research in
Mathematical Sciences for generous hospitality  and the Natural
Sciences and Engineering Research Council of Canada for financial
support in the form of a postdoctoral fellowship.}


\begin{abstract}
We demonstrate that a local mapping $f$ in  a space of bisequences
over $\{0,1\}$ which conserves the number of nonzero sites can be
viewed as a deterministic particle system evolving according to a
local mapping in a space of increasing bisequences over $\ZZ$. We
present an algorithm for determination of the local mapping in the
space of particle coordinates corresponding to the local mapping $f$.
\end{abstract}
\maketitle
\section{Introduction}

Cellular automata (CA) are dynamical systems  characterized by
discreteness in space and time. In general, they can be viewed as cells
in a regular lattice updated synchronously according to a local
interaction rule, where the state of each cell is restricted to a
finite set of allowed values. Among many applications of CA, models of
road traffic flow, first proposed by Nagel and Schreckenberg in 1992
\cite{Nagel92}, attracted substantial attention in recent years. Many
theoretical aspect of the Nagel-Schreckenberg  model are still not
fully understood, and therefore several simplified models have been
proposed, including models based on deterministic cellular automata
\cite{Nagel93,Fukui96c}.

One of the interesting features of these models is the fact that they
can be described using two equivalent paradigms: either as
one-dimensional cellular automata or as systems of interacting
particles on one-dimensional lattice. The simplest example is rule
184, one of the elementary CA rules investigated by Wolfram
\cite{Wolfram94}, and later extensively studied in the context of
surface growth models \cite{Krug88}, as well as in the context of
density classification problem \cite{paper4}. It is one of the only
two (symmetric) non-trivial elementary rules conserving the number of
active sites \cite{paper8,paper15}, and, therefore, can be interpreted
as a rule governing dynamics of particles (cars). Particles (cars)
move to the right        if their right neighbor site is empty, and do
not move if the right neighbor site is occupied, all of them moving
simultaneously at each discrete time step. Using terminology of
lattice stochastic processes, rule 184 can be viewed as a
discrete-time version of totally asymmetric simple exclusion process.

A general question which can be asked is when a given CA rule can be
treated as a rule governing motion of particles, and how to determine
all rules possessing this property. Since the number of particles has
to be conserved, it is clear that all such rules must conserve the
number of nonzero sites. In \cite{paper8,paper15}, CA rules of this
type and their phenomenology were investigated.

In this work, we will formalize concepts introduced in \cite{paper8}.
We will prove that each conservative CA can be associated with a local
mapping of particle coordinates, and we will demonstrate how such a
mapping can be constructed.

\section{Bisequence spaces and their mappings}

The set of definitions given below closely follows terminology used in
\cite{Hedlund68}, with minor modifications.

Let $\CG$ be a countable set with cardinality $g$, which will be
called a {\em symbol set}. If $g$ is finite, we will assume that
$\CG=\{0,1,
\ldots ,g-1\}$, otherwise we will often assume $\CG=\ZZ$.

 A {\em bisequence} over $\CG$ is a function on $\ZZ$ to
$\CG$. Let $X(\CG)$ denote the set of bisequences over $\CG$, i.e.,
$X(\CG)=\CG^\ZZ$. If $x \in X(\CG)$ and $i \in \ZZ$, then $x(i)$ will
be often denoted by $x_i$.

Let $n \in \NN, n>0$. An {\em $n$-block} over $\CG$ is an ordered set
$x_1x_2
\ldots x_n$, where $x_i \in \CG$. The set of all $n$-blocks over $\CG$
will be denoted by $\CB_n(\CG)$.

Let $f$ be a mapping of $\CB_n(\CG)$ into $\CG$. The set of all such
mappings, for a given $n \in \NN, n>0$ will be denoted by $F(\CG,n)$.
Mapping $f$ will be frequently called a {\em local mapping}, or
equivalently a {\em cellular automaton rule}.

Let $f\in F(\CG,n)$ and let $m \in \NN$, $m>0$. Corresponding to $f$
and $m$, we define a mapping $f_{m}:\CB_{m+n-1}(\CG)\rightarrow
\CB_m(\CG)$ as follows. Let $b=b_1 \ldots b_{m+n-1} \in \CB_{m+n-1}(\CG)$
and let $a_i=f(b_i b_{i+1} \ldots b_{i+n-1})$, where $i=1,2,\ldots,m$.
Then we define $f_m(B)=a$, where $a=a_1a_2,\ldots,a_m$.

Finally, let us define a {\em global mapping} $f_{\infty}:X(\CG)
\rightarrow X(\CG)$ corresponding to a local mapping $f\in F(\CG,n)$.
Let $x \in X(\CG)$, and let $y \in X(\CG)$ be defined as
$y_i=f(x_i,x_{i+1},\ldots, x_{i+n-1})$ for all $i\in \ZZ$. Then we define
$f_\infty(x)=y$.

For $a\in \ZZ$ we can define a mapping $f_{\infty,a}:X(\CG)
\rightarrow X(\CG)$, similar to $f_\infty$. If $x \in X(\CG)$, and
if $y \in X(\CG)$ such that $y_i=f(x_{i-a},x_{i+1-a},\ldots,x_{i+n-1-a})$
then, by definition, $f_{\infty,a}=y$. Note that
$f_\infty=f_{\infty,0}$.

As an example, consider a set of mappings $F(\{0,1\},3)$, which were
studied in \cite{Wolfram94} and called elementary cellular automata
rules. There is $2^{2^3}=256$ such mappings, and it is customary,
following \cite{Wolfram94}, to assign them  {\em code numbers} $C(f)$
defined as
\begin{equation} \label{code3}
C(f)=\sum_{x_1,x_2,x_3=0}^{1}f(x_1,x_2,x_3)2^{(2^2x_1+2^1x_2+2^0x_3)}.
\end{equation}
\begin{example}
For instance, the local mapping with code number $184$, often referred
to as simply {\em rule 184}, is defined by
\begin{eqnarray} \label{r184-01}
& & f(0,0,0)=0,\,  f(0,0,1) = 0,\,  f(0,1,0) = 0,\, f(0,1,1)=1 \\ & &
f(1,0,0)= 1,\, f(1,0,1) = 1,\, f(1,1,0)= 0,\, f(1,1,1) = 1.  \nonumber
\end{eqnarray}
It is straightforward to show that the above definition can be written
in a more compact form as
\begin{equation} \label{r184comp}
f(x_1,x_2,x_3)=x_2 + \min\{x_1,1-x_2\} -
\min\{x_2,1-x_3\}.
\end{equation}
Having the local mapping, we can compute $m$-block mappings $f_m$.
For example, $f_3(10100)=101$ because $f(1,0,1)=1$, $f(0,1,0)=0$, and
$f(1,0,0)=1$.
\end{example}

It is a common practice to view CA as dynamical systems and to
investigate trajectories of points in the bisequence space $X(\CG)$,
where by a trajectory of a point $x\in \CG$ we mean the sequence
$\{f^k_\infty(x)\}_{k=0}^{\infty}$,
 superscript $k$ denoting multiple composition
\begin{equation}
f_\infty^k=\underbrace{f_\infty (f_\infty
 ( \cdots  f_\infty(x)))}_{\mbox{$k$ times}}.
\end{equation}

Since CA are often used as model of real physical systems of a finite
size, periodic boundary conditions are frequently employed. Using the
formalism introduced in this section, periodic boundary conditions can
be easily incorporated by assuming that the bisequence under
consideration is periodic with some period $L$, i.e., there exist some
$L \in \NN$ such that $\forall {i\in \ZZ}: x_{i+L}=x_i$. Obviously,
all bisequences belonging to the trajectory of a periodic bisequence
are periodic.

\section{Conservative rules}

As in any other dynamical system, symmetry and conservation laws play
an important role in cellular automata. Additive invariants in
one-dimensional CA have been studied by Hattori and Takesue
\cite{Hattori91,Takesue91}. They obtained conditions which guarantee
the existence of additive conserved quantities, and produced a table
of additive invariants for Wolfram's elementary CA rules. In this
work, we will consider simplest possible additive invariants, namely
the number of active sites (``active'' meaning non-zero).

Let us start from a simple example, rule 184 introduced in the
previous section. Let us consider an $L$-periodic bisequence $x\in
X(\{0,1\})$ and its image under the local mapping with code number
184, as defined by eq. (\ref{r184comp}), $y=f_\infty(x)$. The number
of active sites in a single period of $x$ (to be  referred to as
simply the {\em number of active sites in $x$}) is equal to
$\sum_{i=1}^{L} x_i$. The number of active sites in $y$ is, using eq.
(\ref{r184comp}),
\begin{eqnarray*}
\sum_{i=1}^{i=L} y_i =\sum_{i=1}^{i=L} f(x_{i+1},x_{i+2},x_{i+3})=
\sum_{i=1}^{i=L} x_{i+2} +       \\
\sum_{i=1}^{i=L}\min\{x_{i+1},1-x_{i+2}\} -
\sum_{i=1}^{i=L}\min\{x_{i+2},1-x_{i+3}\}.
\end{eqnarray*}
Since for a periodic lattice the last two sums cancel, we obtain
\begin{equation}
\sum_{i=1}^{i=L} x_i=\sum_{i=1}^{i=L} y_i,
\end{equation}
meaning that the number of active sites is conserved.

In general, a local mapping $f \in F(\CG,n)$ will be called {\em
$L$-conservative} if for any $L$-periodic bisequence $x\in
X({\mathcal{G}})$ the following condition is satisfied:
\begin{equation}
\sum_{i=1}^{i=L} f(x_{i}, x_{i+1}, \ldots , x_{i+n-1}) =
\sum_{i=1}^{i=L} x_i.
\end{equation}
Using the fact that $x$ is periodic, this condition can be rewritten
as
\begin{eqnarray}
f(x_1,x_2,\ldots,x_n) + f(x_2,x_3,\ldots,x_{n+1}) + \cdots  +
f(x_L,x_1,\ldots,x_{n-1})  \nonumber \\
\qquad= x_1+x_2+\cdots+x_L.
\label{L-cond}
\end{eqnarray}
If for every $L>0$ a mapping $f \in F(\CG,n)$ is $L$-conservative,
than it is said to be {\em conservative}. The following theorem
\cite{paper8} is helpful in determining if a given rule is
conservative.

\begin{theorem}[\cite{paper8}]
If a mapping $f \in F(\CG,n)$ is $L$-conservative for $L=2n-2$, then
it is conservative.
\end{theorem}
\begin{proof}
To prove the above result, we shall show that, if $L>2n-2$, any
equation  of (\ref{L-cond}) type, i.e., equation expressing
conservation condition for $L$-periodic configuration, is a linear
combination of three equations which express, respectively,
conservation conditions for $(L-1)$-, $(2n-3)$-, and $(2n-2)$-periodic
sequences. More precisely, for all $L$-periodic bisequences with period
$L$, $\{x_1,x_2,\ldots,x_L\}$, Equation~(\ref{L-cond}) can be written
\begin{eqnarray}
 & &\Big(f(x_1,x_2,\ldots,x_n) + f(x_2,x_3,\ldots,x_{n+1}) + \cdots
 + f(x_{L-1},x_1,\ldots,x_{n-1})\Big) \nonumber \\
 & & -\Big(f(x_1,x_2,\ldots,x_{n-2},x_{L-n+1},x_{L-n+2})\nonumber \\
 & & + f(x_2,x_3,\ldots,x_{L-n+3}) + \cdots +
 f(x_{n-2},x_{L-n+1},\ldots,x_{L-1})\nonumber \\
 & & + f(x_{L-n+1},x_{L-n+2},\ldots,x_{L-1},x_1) + \ldots +
 f(x_{L-1},x_1,\ldots,x_{n-2},x_{L-n+1})\Big)\nonumber \\
 & & +\Big(f(x_1,x_2,\ldots,x_{n-2},x_{L-n+1},x_{L-n+2})+
 f(x_2,x_3,\ldots,x_{L-n+3}) + \cdots \nonumber \\
 & &+
 f(x_{n-2},x_{L-n+1},\ldots,x_{L-1}) +
 f(x_{L-n+1},x_{L-n+2},\ldots,x_{L-1},x_L) + \ldots\nonumber \\
 & &+
 f(x_{L-1},x_L,x_1,\ldots,x_{n-2}) +
 f(x_L,x_1,\ldots,x_{n-2},x_{L-n+1})\Big)\nonumber \\
 & & =
 (x_1+x_2+\cdots+x_{L-1})
 - (x_1+\cdots+x_{n-2}+x_{L-n+1}+\cdots+x_{L-1})\nonumber \\
 & &+ (x_1+\cdots+x_{n-2}+x_{L-n+1}+\cdots+x_L).
\label{eqn-decomp}
\end{eqnarray}

The  above rearrangement of terms\footnote{As pointed out by a
referee, this rearrangement of terms can be also understood as a
representation of an $L$-cycle in a de Bruijn graph as the sum of an
$L-1$-cycle, a $2n-2$-cycle, and a $2n-3$-cycle.} is valid if
$x_{L-1}=x_L$, and for almost every periodic bisequence, we can choose
a ``coordinate system'' (labeling of sites) in which it is true. The
only exception is, for even $L$, the bisequence with period
$1010\ldots 10$. Verifying (\ref{eqn-decomp}) is in this case
possible, but needs to be done with a different method (see below). By
induction, relation (\ref{eqn-decomp}) shows that the conservation
condition for $L$-periodic bisequence is a linear combination of
conservation conditions for  $(2n-3)$- and $(2n-2)$-periodic
bisequences. Using similar rearrangement of terms as in
(\ref{eqn-decomp}), it can be shown that conservation conditions for
$(2n-2)$-periodic bisequences imply conservation conditions for
$(2n-2)$-periodic bisequences, concluding the proof.

Consider now the exception mentioned  in the preceding paragraph, the
bisequence $1010\ldots 10$. Conservation condition for arbitrary $L$
reads
\begin{equation} \label{per1}
\underbrace{f(1010\ldots 10)+f(0101\ldots 01)+\cdots+f(0101\ldots 01)}_L
= \frac{L}{2},
\end{equation}
if $n$ is even, and
\begin{equation} \label{per2}
\underbrace{f(1010\ldots 01)+f(0101\ldots 10)+\cdots+f(0101\ldots 10)}_L
= \frac{L}{2},
\end{equation}
if $n$ is odd.
That is,
\begin{equation}
f(1010\ldots 10)+f(0101\ldots 01) = 1,
\end{equation}
if $n$ is even, and
\begin{equation}
f(1010\ldots 01)+f(0101\ldots 10) = 1,
\end{equation}
if $n$ is odd. In the above two equations $L$ does not occur at all,
so obviously, if eq. (\ref{per1}) or (\ref{per2}) is true for
$L=2n-2$, it is true for any $L$.
\end{proof}

Using this theorem, if one wants to find all mappings $f \in F(\CG,n)$
for some $n$ which are conservative, it is enough to check which of
them are $2n-2$-conservative. In fact, for many mappings it is
possible to show that they are {\em not} conservative (and thus
eliminate them from a list of mappings ``suspected'' for being
conservative) by employing some of their elementary properties.
\begin{theorem}
Let $n \in \NN$ and $n>0$. If a mapping $f \in F(\CG,n)$ is
conservative, then
\begin{itemize}
\item[(a)] $f(0,0,\ldots,0)=0$,
\item[(b)] $f(1,1,\ldots,1)=1$,
\item[(c)]
$\displaystyle \sum_{x_1,x_2, \ldots, x_n \in\{ 0,1\}}  f(x_1,x_2,
\ldots, x_n) = 2^{n-1}$, i.e., $f$ is $1$-balanced.
\end{itemize}
\end{theorem}
\begin{proof} Part (a) becomes obvious if we consider periodic
bisequence consisting of all zeros, i.e., containing no active sites.
Image of this configuration must also contain no active sites, and it
is only possible if $f(0,0,\ldots,0)=0$. Proof of (b) is identical.

Part (c) says that among all possible configurations of arguments
$x_1,x_2, \ldots, x_n$ of $f$ (there are $2^n$ such configurations),
exactly half results in $f(x_1,x_2, \ldots, x_n)=1$, as can be seen,
for example, in definition of rule 184 (eq. \ref{r184-01}).

To prove part (c), consider periodic bisequence $t\in X(\{0,1\})$ of
period $m=2^n$. Now, let us construct a set of $m$ blocks
$A=\{b^{(j)}\}_{j=1}^m$ such that $b^{(j)}=t_j,t_{j+1},\ldots,
t_{j+n-1}$ (superscript $(j)$ denotes here just a consecutive number
of the block $b^{(j)}$ in the set $A$).

Assume that we can find $t$ such that all blocks in $A$ are different.
This means that each possible block of length $n$ occurs in $A$ once
and only once, and therefore
\begin{equation} \label{t1}
\displaystyle \sum_{x_1,x_2, \ldots, x_n \in\{ 0,1\}}  f(x_1,x_2,
\ldots, x_n) = \sum_{i=1}^m f(t_i,t_{i+1},\ldots, t_{i+n-1})=N_{1},
\end{equation}
where $N_{1}$ is the number of 1's in a single period of $t$.

On the other hand, $\bar{t}\in X(\{0,1\})$, which is obtained from $t$
by replacing all zeros by ones and {\em vice versa}, must also have
the same property as $t$, i.e., in a set $\bar{A}=\{b^{(j)}\}_{j=1}^m$
such that $b^{(i)}=\bar{t}_i,\bar{t}_{i+1},\ldots, \bar{t}_{i+n-1}$,
all elements are different, hence
\begin{equation} \label{t2}
\displaystyle \sum_{x_1,x_2, \ldots, x_n \in\{ 0,1\}}  f(x_1,x_2,
\ldots, x_n) = \sum_{i=1}^m f(\bar{t}_i,\bar{t}_{i+1},\ldots,
\bar{t}_{i+n-1})=N_{0},
\end{equation}
where $N_{0}$ is the number of 0's in a single period of $t$.
Comparing (\ref{t1}) and (\ref{t2}) we obtain $N_0=N_1=2^n/2=2^{n-1}$,
exactly as required.

The only problem left is to show that, indeed, for any $n>0$, we can
construct $2^n$-periodic configuration $t$ such that all blocks
occurring in a single period of $t$ are different (and therefore,
constitute a set of all possible blocks of length $n$). For example,
for $n=2$, consider $t$ with a period $1100$. One can easily check
that all possible blocks of length $2$ occur in a single period of
$t$: $11$, $10$, $00$ and $01$. For $n=3$,  $t$   with a period
$11101000$ has the same property. Again, one can see that blocks
occurring in the period of $t$, $111$, $110$, $101$, $010$, $100$,
$000$, $001$, and $011$, are all possible blocks of length $3$.

For a general $n$, the required $t$ is equivalent to a hamiltonian
cycle in the de Bruijn graph  \cite{DeBruij46} of dimension $n$ (or an
eulerian cycle in the de Bruijn graph of dimension $n-1$). It has been
demonstrated that such a cycle always exists (more precisely, for a
given $n$, there exist exactly $2^{2^{n-1}-n}$ of such cycles -- see,
for example, review article \cite{Ralston}).
\end{proof}

\section{Conservative cellular automata viewed as deterministic
particle systems}

Since conservative rules conserve the number of active sites, one can
identify active sites with ``particles'' which change position after
each application of the rule $f_\infty$, but their number does not
change. Therefore, instead of describing this system of particles in
terms of lattice sites being occupied or empty, we can describe it by
specifying list of coordinates of all particles.

\begin{definition}
An {\em increasing bisequence} over $\ZZ$ is an increasing function on
$\ZZ$ to $\ZZ$. Set of all increasing sequences over $\ZZ$ will be
denoted by $\tilde{X}(\ZZ)$. If $s \in \tilde{X}(\ZZ)$ and $i
\in \ZZ$, then $s(i)$ will be often denoted by $s_i$. Now, let us define
a mapping $\phi: \tilde{X}(\ZZ) \mapsto X(\{0,1\})$ as follows: let $s
\in
\tilde{X}(\ZZ)$ be an increasing bisequence and let $x
\in X(\{0,1\})$ be defined by
\begin{equation}
x_i= \left\{ \begin{array}{ll}
 1,    & \mbox{if $\exists k \in \ZZ:  s_k=i$}, \\
 0,    & \mbox{otherwise},
\end{array}
\right.
\end{equation}
for every $i \in \ZZ$. Then, by definition, $\phi(s)=x$.
\end{definition}

Using the notion of particles, $s$ in the above definition can be
understood as a list of coordinates of particles on one-dimensional
lattice. We require this list to be increasing, so no two particles
occupy the same position. On the other hand, $x\in X(\{0,1\})$ is just
a list of occupancy numbers: for all sites, $x_i$ is either $1$
(meaning site $i$ is occupied) or $0$ (empty site). Transformation
$\phi$ takes a list of particle coordinates $s$ and returns the
corresponding list of occupancy numbers. For example, if in some $s\in
\tilde{X}(\ZZ)$ we have $s_1=7$, $s_2=9$, $s_3=12$, then in the corresponding
$x=\phi(s)$, sites $7$, $9$, and $12$ are occupied, meaning that
$x_7=1$, $x_9=1$, and $x_{12}=1$. Note that since $s$ is an increasing
bisequence, we can immediately conclude that the site between $7$ and
$9$ as well as both sites between $9$ and $12$ must be empty, i.e.,
$x_8=0$ and $x_{10}=x_{11}=0$.

Since conservative mappings conserve the number of particles, we would
now like to find a local mapping which transforms coordinate lists
$y\in \tilde{X}(\ZZ)$ with the same effect as conservative local
mappings $f\in (\{0,1\}),n)$ transform occupancy lists $x\in
X(\{0,1\})$.

\begin{theorem} \label{theofg}
Let $n\in N$, and let $f\in F(\{0,1\},n)$. There exist $m\in \NN$, $a
\in \ZZ$, and $g \in F(\ZZ,m)$ such that
\begin{itemize}
\item[(i)] for every $s\in \tilde{X}(\ZZ)$, $g_{\infty,a}(x) \in \tilde{X}(\ZZ)$,
\item[(ii)] $\phi g_{\infty,a} =f_\infty \phi$.
\end{itemize}
\end{theorem}
Note that Theorem \ref{theofg} can be represented as
\begin{equation}
 \begin{CD}
 \tilde{X}(\ZZ) @>{\phi}>> X(\{0,1\}) \\
 @VV{g_{\infty,a}}V  @VV{f_\infty}V\\
  \tilde{X}(\ZZ) @>{\phi}>>  X(\{0,1\}). \\
 \end{CD}
\end{equation}
Before we present the proof, let us consider an example which will
clarify the meaning of Theorem \ref{theofg}.
\begin{example}
Equation \ref{r184comp} defined a local function for cellular
automaton rule 184 as
\begin{equation} \label{fdef184}
f(x_1,x_2,x_3)=x_2 + \min\{x_1,1-x_2\} -
\min\{x_2,1-x_3\}.
\end{equation}
Corresponding rule $g$ of Theorem \ref{theofg} (constructed by an
algorithm to be presented later) is defined as:
\begin{equation} \label{gdef184}
g(s_1,s_2)=s_1 + \min \{s_2-s_1-1,1\}.
\end{equation}
Consider now a bisequence $s \in \tilde{X}(\ZZ)$, in which $s_1=1$,
$s_2=2$, and $s_3=4$, while $s_0<1$, $s_4>7$. (this means that among
lattice sites $i=1
\ldots 8$ only $i=1,2$ and $4$ are occupied). Let $p=g_\infty(s)$,
which means that $p_i=g(s_i,s_{i+1})$, yielding $s_1=1$, $s_2=3$, and
$s_3=5$. List  of particle coordinates $\{\ldots 1,2,4,\ldots\}$ is
transformed by $g_\infty$ into $\{\ldots 1,3,5,\ldots\}$

Now, consider $x\in X(\{0,1\})$ such that $x=\phi(s)$, and let us find
$y=f_\infty(x)$, so
 that\footnote{Strictly speaking, we are using
 $f_{\infty,1}$ here, not $f_\infty$, mainly due to historical
 convention commonly employed for elementary ($3$-input) cellular
 automata. If we wanted to use $y_i=f(x_{i},x_{i+1},x_{i+2})$,
 we would have to slightly redefine $g$, so that $g(s_1,s_2)=s_1-1 + \min \{s_2-s_1-1,1\}$.
 However, the form used in the example is more intuitive (it represents
 asymmetric exclusion process with discrete time).
  }
$y_i=f(x_{i-1},x_i,x_{i+1})$. Both $x$ and $y$ can be represented as
\begin{center}
\begin{tabular}{|c|c|c|c|c|c|c|c|c|} \hline
$\cdots$&1&1&0&1&0&0&0&$\cdots$\\ \hline
$\cdots$&1&0&1&0&1&0&0&$\cdots$\\ \hline
\end{tabular}
\end{center}
where the first line corresponds to $x$, and the second line to $y$.
One can readily see that in the first line particles are located at
sites $1,2$ and $4$, while at second line at $1,3$ and $5$, confirming
that $\phi g_\infty =f_\infty \phi$. Note that in this example
application of $g_\infty$ results in a well defined motion of
particles: if the site on the right of a given particle is empty, it
moves to that site, otherwise it stays in the same place. Using
notation introduced in \cite{paper8}, this can be written as
\begin{equation}
\overset{\curvearrowright}{10},\quad
\overset{\circlearrowright}{1}1.
\end{equation}
(arrow shows where the particle will move, circular arrow indicates
that the particle stays in the same place).
\end{example}

\section{Labeling scheme for lattice sites}

In order to construct function $g$ of Theorem \ref{theofg}, we will
first introduce transformation of coordinate space $\tilde{X}(\ZZ)$ to
intermediate (``mixed'') space $\Psi$, which combines both coordinates
of particles and information about occupancy of lattice sites. This can
be done by labeling occupied lattice sites with consecutive
integers, and empty lattice sites  with extra symbol
$\no$, so that configurations of particles become bisequences over $\ZZ
\cup \{\no\}$ (the extra symbol ``$\no$'' had to be introduced since ``0''
can be a particle label, and therefore cannot be used to denote empty
sites).

\begin{definition} \label{psidef}
Let $x\in \tilde{X}(\ZZ)$. Let $y\in X(\ZZ \cup \{\no \})$ be a
bisequence constructed as follows: $y_i=j$ if there exists $j\in \ZZ$
such that $x_j=i$, otherwise $y_i=\no$. By definition, $\psi(x)=y$.
Bisequences $y$ as defined above will be called {\em increasing
labeling bisequences}, and  the set of all such bisequences will be
denoted by $\Psi=\psi(\tilde{X}(\ZZ))$.
\end{definition}

Note that $\psi$ transforms a bisequence of particle coordinates, such
as $x=\ldots 1, 2, 4, \ldots$ into a bisequence $\psi(x)$:
\begin{center}
\begin{tabular}{|c|c|c|c|c|c|c|c|c|} \hline
$\cdots$&1&2&$\no$&3&$\no$&$\no$&$\no$&$\cdots$\\ \hline
\end{tabular}
\end{center}
which resembles $\phi(x)$, except that occupied sites are now labeled
with unique integers, increasing from left to right (the label of an
occupied site is always larger by $1$ than the label of the closest
occupied site on the left). If we define, for $k\in \ZZ \cup \{\no\}$,
\begin{equation}
\|k\|= \left\{ \begin{array}{ll}
 1,    & \mbox{if $k \in \ZZ$}, \\
 0,    & \mbox{if $k=\no$},
\end{array}
\right.
\end{equation}
then, obviously, $(\phi(x))_i =
\|(\psi(x))_i\|$ for every $i\in \ZZ$.

We will now define a mapping which plays a similar role in $\Psi$ as
$f$ plays in $X(\{0,1\})$. Corresponding to $f$, mapping $\hat{f} \in
F(\ZZ \cup \{\no \},n)$ will be defined as follows.

Define
\begin{multline}   \label{defg}
 G(x_1,x_2,\ldots, x_n)= f(0,0,\ldots,0,\|x_1\|) +
f(0,0,\ldots,\|x_1\|,\|x_2\|) +\\
\ldots + f(0,\|x_1\|,\|x_2\|, \ldots, \|x_n\|) +
 f(\|x_1\|,\|x_2\|,\ldots, \|x_n\|),
\end{multline}
and
\begin{equation} \label{defh}
H(x_1,x_2,\ldots, x_n, k)=\min\{i\in \NN: \sum_{j=1}^{i}\|x_j\| = k\}.
\end{equation}
Now, function $\hat{f}$ is defined as

 \begin{equation} \label{defhat}
\hat{f}(x_1,x_2, \ldots, x_n)= \left\{ \begin{array}{ll}
 x_l,    & \mbox{if $f(\|x_1\|,\|x_2\|, \ldots, \|x_n\|)=1$}, \\
 \no,    & \mbox{otherwise},
\end{array}
\right.
\end{equation}
where $l=H(x_1,x_2,\ldots, x_n, G(x_1,x_2,\ldots, x_n))$.

\begin{example}
For rule $184$, as defined in (\ref{r184comp}), applying the above
definition we obtain for every $x_1, x_2, x_3 \in \ZZ$,
\begin{align*}
 \hat{f}(\no,\no,\no)      =\no,
 \hat{f}(\no,\no,x_1)    =\no,
 \hat{f}(\no,x_1,\no)    =\no,
 \hat{f}(\no,x_1,x_2)  =x_1, \\
 \hat{f}(x_1,\no,\no)    =x_1,
 \hat{f}(x_1,\no,x_2)  =x_1,
 \hat{f}(x_1,x_2,\no)  =\no,
 \hat{f}(x_1,x_2,x_3)=x_2.
\end{align*}
\end{example}

An immediate consequences of the definition of $\hat{f}$ is
$\|\hat{f}(x_1, x_2, \ldots, x_n)\|=f(\|x_1,\|,
\|x_2,\|,\ldots,\|x_n,\|)$, and therefore:
\begin{lemma}
Let $n\in \NN$, $n>0$,  and let $f\in F(\{0,1\},n)$. Then for every
$x\in \Psi$, $\|\hat{f}_\infty(x)\| = f_\infty(\|x\|)$.
\end{lemma}
A crucial property of $\hat{f}$ is that an image of an increasing
labeling bisequence under $\hat{f}_\infty$ is also an increasing
labeling bisequence:

\begin{theorem} \label{theomixed}
Let $n\in \NN$, $n>0$, and let $f\in F(\{0,1\},n)$. If $x\in \Psi$,
then $\hat{f}_\infty(x) \in \Psi$.
\end{theorem}

\begin{proof}
Let us consider an increasing labeling bisequence $x$ and its image
$y=\hat{f}_\infty(x)$. Consider two sites in $y$, $y_i$ and $y_j$,
$i<j$, such that all sites between them are empty, i.e. $y_k=\no$ for
all $i<k<j$. We want to show that $y_j=y_i+1$, since this would
demonstrate that $y$ is indeed an increasing labeling bisequence.
Obviously, $y_i=\hat{f}(x_i, x_{i+1}, \ldots, x_{i+n-1})$ and
$y_j=\hat{f}(x_j, x_{j+1}, \ldots, x_{j+n-1})$. There must be some
$p,q$ such that $y_i=x_p$ and $y_j=x_q$, which, according to the
definition of $\hat{f}$, must satisfy
\begin{align}
 p=&i-1+H(x_i,x_{i+1}, \ldots, x_{i+n-1},G(x_i,x_{i+1}, \ldots, x_{i+n-1})), \\
 q=&j-1+H(x_j,x_{j+1}, \ldots, x_{j+n-1},G(x_j,x_{j+1}, \ldots, x_{j+n-1})). \nonumber
\end{align}
As a first step of the proof, we will  find relationship between
$G(x_i,x_{i+1}, \ldots, x_{i+n-1}))$ and $G(x_j,x_{j+1}, \ldots,
x_{j+n-1}))$.

Let us denote the number of nonzero sites in the block $x_i,
x_{i+1},\ldots, x_{j-1}$ by $N$, i.e., $N=\|x_i\|+\|x_{i+1}\|+,
\ldots, +\|x_{j-1}\|$. Consider now a bisequence $t\in \Psi$ such that
$t_l=x_l$ for $i \leq l \leq j+n-1$, and $t_l=\no$ otherwise, and
another one, $u\in \Psi$, such that $u_l=x_l$ for $j \leq l \leq
j+n-1$, and $t_l=\no$ otherwise. Due to the fact that $f$ is
conservative, we have
\begin{equation} \label{turel}
\sum_{l=-\infty}^{\infty} \|\hat{f}(t_l,t_{l+1}, \ldots, t_{l+n-1})\|=N+
\sum_{l=-\infty}^{\infty} \| \hat{f}(u_l,u_{l+1}, \ldots, u_{l+n-1})\|.
\end{equation}
Using definitions of $t$ and $u$, and the fact that
$f(0,0,\ldots,0)=0$, equation (\ref{turel}) becomes
\begin{multline}
 \|\hat{f}(\no,\ldots,\no, x_i)\| +
 \|\hat{f}(\no,\ldots,\no,x_i, x_{i+1})\| + \ldots +
 \|\hat{f}(x_i,x_{i+1},\ldots,x_{i+n-1})\|+ \\
 \|\hat{f}(x_{i+1},x_{i+2},\ldots, x_{i+n})\| + \ldots +
 \|\hat{f}(x_j, x_{j+1},\ldots,x_{j+n-1})\|+ \\
 \|\hat{f}(x_{j+1},x_{j+2},\ldots,\no)\|+ \ldots +
 \|\hat{f}(x_{j+n-1},\no,\ldots,\no)\|=N+\\
 \|\hat{f}(\no,\ldots,\no, x_j)\| +
 \|\hat{f}(\no,\ldots,\no,x_j, x_{j+1})\| + \ldots +
 \|\hat{f}(x_j,x_{j+1},\ldots,x_{j+n-1})\|+ \\
 \|\hat{f}(x_{j+1},x_{j+2},\ldots,\no)\|+ \ldots +
 \|\hat{f}(x_{j+n-1},\no,\ldots,\no)\|.
\end{multline}
However, we know that there are no particles between $y_i$ and $y_j$,
therefore
\begin{equation}
 \|\hat{f}(x_{l}, x_{l+1}, \ldots, x_{l+n-1})\|=0
\end{equation}
for all $l$ such that $i<l<j$. Taking this into account, and using
definition of $G$ (eq. \ref{defg}) we obtain, after some
cancellations,
\begin{multline}
G(x_i,x_{i+1}, \ldots, x_{i+n-1})+\|f(x_j,x_{j+1},
\ldots, x_{j+n-1})\|=\\N+G(x_j,x_{j+1},
\ldots, x_{j+n-1}).
\end{multline}
Since $\|f(x_j,x_{j+1}, \ldots, x_{j+n-1})\|=1$, this finally becomes
\begin{equation}
G(x_i,x_{i+1}, \ldots, x_{i+n-1})=N-1+G(x_j,x_{j+1},
\ldots, x_{j+n-1}).
\end{equation}

Having this relationship, note that, according to definition of $H$
(eq. \ref{defh}), $x_p$ is $k_p$-th particle in the block
$x_i,x_{i+1},\ldots,x_{i+n-1}$ starting from the left, where
$k_p=G(x_i,x_{i+1},\ldots, x_{i+n-1})$. Obviously, it is also  is
$k_p$-th particle in the block $x_i,x_{i+1},\ldots,x_{j+n-1}$ starting
from the left.

Similarly, $x_q$ is $k_q$-th particle in the block $x_j,x_{j+1},
\ldots,x_{j+n-1}$, where $k_q=G(x_j,x_{j+1},\ldots, x_{j+n-1})$.
Knowing that sites $x_i,\ldots,x_{j-1}$ contain exactly $N$ particles,
we conclude that $x_q$ must also be $(N+k_q)$-th particle in the block
$x_i,x_{i+1},\ldots,$ $x_{j+n-1}$. However, $k_q=k_p-N+1$, therefore:

\vspace{0.5em}
{\em \noindent Particle $x_p$ is $k_p$-th particle from the left in
the block $x_i,x_{i+1}, \ldots, x_{j+n-1}$, while $x_q$ is
$(k_p+1)$-th particle in the same block.}
\vspace{0.5em}

\noindent This means that there is no other particle between $x_p$ and $x_q$,
hence $x_q=x_p+1$ (we assumed that $x$ is an increasing labeling bisequence),
and, finally, $y_j=y_i+1$, which is exactly what
what we wanted to show.
\end{proof}

\section{Construction of local mapping in $\tilde{X}(\ZZ)$}
We are now ready to see how $g$ of Theorem \ref{theofg} can be
constructed.

First, let us extend definition of the mapping $\psi$ (Definition
\ref{psidef}) to blocks of particle coordinates. Let $b\in \CB_m(\ZZ)$
be called an {\em increasing block} if $b_1<b_2<\ldots<b_m$. Set of
all such blocks will be denoted by $\tilde{\CB}_m(\ZZ)$. For $m>1$, $b
\in \tilde{\CB}_m(\ZZ)$ let us define $c=\psi(b)$ such that
$c\in \CB_{b_m-b_1+1}(\ZZ \cup \{\no\})$ and that, for every $l\in
\NN$  satisfying $1\leq l \leq b_m-b_1+1$,  if there exist $k\in \NN$
such that $b_k=l$, then $c_l=k-b_1+1$, otherwise $c_l=\no$. Mapping
$\psi$, as before, transforms finite and increasing list of particle
coordinates into finite block in particles  are located at sites
$b_l-b_1+ 1$ and labeled with labels $l$. For example, if
$b=\{9,10,14\}$, then $\psi(b)=\{1,2,\no,\no,3\}$.

Similarly as we defined increasing labeling bisequences, we define
{\em blocks with increasing labels} as elements of the set
$\psi(\tilde{\CB}_m(\ZZ))$. This set will be denoted by
$\tilde{\CB}_m(\ZZ \cup \{\no\})$

In Theorem \ref{theomixed}, we proved that $\hat{f}_\infty$ maps
increasing labeling bisequences to increasing labeling bisequences.
The same is true for blocks with increasing labels (and can be proved
using almost identical reasoning as in the proof of Theorem
\ref{theomixed}, therefore we omit the proof):
\begin{corollary}
For any $m,n\in N$, $m\geq n$, and for $f \in F(\{0,1\},n)$, if $b\in
\tilde{\CB}_m(\ZZ \cup
\{\no\})$, then $\hat{f}_{m}(b)
\in \tilde{\CB}_{m-n+1}(\ZZ \cup \{\no\})$.
\end{corollary}

Let us now assume that we have a list of particle coordinates $s\in
\tilde{X}(\ZZ)$, and we want to find new list of particles'
positions $t$ such that $\phi(t)=f_\infty(\phi(s))$, where $f\in
F(\{0,1\},n)$ for some $n\in \NN$, $n>1$. Using $\hat{f}$, this can be
done as follows:
\begin{algorithm} \label{algog}
For each particle coordinate $s_i$ perform the following
 steps:
\begin{enumerate}
\item construct a block $b=\{s_{i-n+1},s_{i-n+2},
\ldots, s_{i+n-1}\}$
\item find block $c=\hat{f}_{s_{i+n-1}-s_{i-n+1}+1}(\psi(b))$, this
 will be a block with increasing labels, with particles in their new
 positions
\item find $d=\psi^{-1}(c)$, this will be a block containing new positions
of particles
\item find new coordinate of particle $i$, which is simply equal to $d_i$
\end{enumerate}
\end{algorithm}
Note that the size of the block $b$ is chosen large enough to ensure
that particle with label $i$ remains in the block after step $2$. For
many rules, smaller neighborhood will suffice.

The above algorithm provides  a way to compute $t_i$ given $s_i$ and
coordinates of particles in the neighborhood of $s_i$. This means
that it defines a local mapping $g$ requested by theorem Theorem
\ref{theofg}. In fact, it is possible to write explicit expression for
$g$ based on the Algorithm \ref{algog}. Corresponding to $f\in
F(\{0,1\},n)$, mapping $g: \CB_{2n-1}(\ZZ) \rightarrow \ZZ$ is defined
as follows:
\begin{equation}
g(b)= (\psi^{-1}(\hat{f}_{}(\psi(b))))_n,
\end{equation}
and parameter $a$ of Theorem \ref{theofg} equals to $n$.

As a final note, let us remark that the relationship of $f$,
$\hat{f}$, and $g$ can be represented as

\begin{equation}
 \begin{CD}
 \tilde{X}(\ZZ) @>{\psi}>> \Psi @>{\|\cdot \|}>> X(\{0,1\}) \\
 @VV{g_{\infty,a}}V @VV{\hat{f}_{\infty}}V @VV{f_\infty}V\\
 \tilde{X}(\ZZ) @>{\psi}>> \Psi @>{\|\cdot \|}>> X(\{0,1\}). \\
 \end{CD}
\end{equation}

We should also mention that $g$ constructed here is not the only local
mapping satisfying Theorem \ref{theomixed}. Infinite number of local
mappings with the same properties as $g$ can be constructed as a
superposition of $g$ and the shift map on $\tilde{X}(\ZZ)$ (by the
shift map we mean $\sigma:\tilde{X}(\ZZ) \rightarrow
\tilde{X}(\ZZ)$ such that $(\sigma(x))_i=x_{i-1}$ for every
$x\in \tilde{X}(\ZZ)$ and every $i\in \ZZ$). All mappings $\sigma^kg$
as well as $\sigma^{-k}g$ for any $k\in \NN$ could be used in place of
$g$. As a convention, we will always use a local mapping such that if
$x_i=i$ for every $i \in \ZZ$, and $y=g_\infty(x)$, then $y_i=i$,
i.e., if all lattice sites are occupied in $x$, then in $y$ they are
in the same position.

\section{Example: four-input conservative rules}

Among $2^{2^4}$ local mappings $f \in F(\{0,1\},4)$ (four-input
rules), only 22 rules are conservative \cite{paper8}. For all of them,
we computed $\hat{f}$, as shown in Table 1.

\begin{table}
\caption{Four-input rules conserving density of nonzero sites.}
\begin{center}
\begin{tabular}{|c|c|l|} \hline
$C(f)$ & Binary form     & $\hat{f}$                                                  \\ \hline
 43690 & 1010101010101010 & $x_4\no x_4\no x_4\no x_4\no x_4\no x_4\no x_4\no x_4\no $ \\
 43944 & 1010101110101000 & $x_3\no x_2\no x_3\no x_1x_1x_3\no x_2\no x_3\no \no \no $ \\
 47288 & 1011100010111000 & $x_3\no x_2x_2x_3\no \no \no x_3\no x_2x_2x_3\no \no \no $ \\
 48268 & 1011110010001100 & $x_3\no x_2x_2x_3x_3\no \no x_3\no \no \no x_3x_3\no \no $ \\
 48770 & 1011111010000010 & $x_3\no x_2x_2x_3x_3x_4\no x_3\no \no \no \no \no x_4\no $ \\
 49024 & 1011111110000000 & $x_2\no x_1x_1x_1x_1x_1x_1x_2\no \no \no \no \no \no \no $ \\
 51448 & 1100100011111000 & $x_3x_3\no \no x_3\no \no \no x_3x_3x_2x_2x_3\no \no \no $ \\
 52428 & 1100110011001100 & $x_3x_3\no \no x_3x_3\no \no x_3x_3\no \no x_3x_3\no \no $ \\
 52930 & 1100111011000010 & $x_3x_3\no \no x_3x_3x_4\no x_3x_3\no \no \no \no x_4\no $ \\
 53184 & 1100111111000000 & $x_2x_2\no \no x_1x_1x_1x_1x_2x_2\no \no \no \no \no \no $ \\
 56528 & 1101110011010000 & $x_2x_2\no x_2x_1x_1\no \no x_2x_2\no x_2\no \no \no \no $ \\
 57580 & 1110000011101100 & $x_3x_3x_4\no \no \no \no \no x_3x_3x_4\no x_3x_3\no \no $ \\
 58082 & 1110001011100010 & $x_3x_3x_4\no \no \no x_4\no x_3x_3x_4\no \no \no x_4\no $ \\
 58336 & 1110001111100000 & $x_2x_2x_2\no \no \no x_1x_1x_2x_2x_2\no \no \no \no \no $ \\
 59946 & 1110101000101010 & $x_3x_3x_4\no x_4\no x_4\no \no \no x_4\no x_4\no x_4\no $ \\
 60200 & 1110101100101000 & $x_2x_2x_2\no x_3\no x_1x_1\no \no x_2\no x_3\no \no \no $ \\
 61680 & 1111000011110000 & $x_2x_2x_2x_2\no \no \no \no x_2x_2x_2x_2\no \no \no \no $ \\
 62660 & 1111010011000100 & $x_2x_2x_2x_2\no x_3\no \no x_2x_2\no \no \no x_3\no \no $ \\
 63544 & 1111100000111000 & $x_2x_2x_2x_2x_3\no \no \no \no \no x_2x_2x_3\no \no \no $ \\
 64524 & 1111110000001100 & $x_2x_2x_2x_2x_3x_3\no \no \no \no \no \no x_3x_3\no \no $ \\
 65026 & 1111111000000010 & $x_2x_2x_2x_2x_3x_3x_4\no \no \no \no \no \no \no x_4\no $ \\
 65280 & 1111111100000000 & $x_1x_1x_1x_1x_1x_1x_1x_1\no \no \no \no \no \no \no \no $ \\  \hline
\end{tabular}
\end{center}
\end{table}
The first column in this
table represents code number $C(f)$, defined similarly as
(\ref{code3}):
\begin{equation} \label{code4}
C(f)=\sum_{x_1,x_2,x_3,x_4=0}^{1}f(x_1,x_2,x_3,x_4)
2^{(2^3x_1+2^2x_2+2^1x_3+2^0x_4)}.
\end{equation}
The second column is a binary representation of $C(f)$, meaning that
it is a sequence of $16$ binary digits $a_{15}a_{14} \ldots a_0$ such
that for every $x_1,x_2,x_3,x_4 \in \{0,1\}$
\begin{equation}
 a_{2^3x_1+2^2x_2+2^1x_3+2^0x_4}=f(x_1,x_2,x_3,x_4).
\end{equation}
Entries in the third column are constructed in a similar way. They are
sequences $b_{15} b_{14} \ldots b_0$ such that if in the definition of
$\hat{f}$ we have $\hat{f}(x_1,x_2,x_3,x_4) =x_j$, then we define
\begin{equation}
b_{2^3\|x_1\|+2^2\|x_2\|+2^1\|x_3\|+2^0\|x_4\|}=x_j,
\end{equation}
and if
\begin{equation}
\hat{f}(x_1,x_2,x_3,x_4) =\no,
\end{equation}
then
\begin{equation}
b_{2^3\|x_1\|+2^2\|x_2\|+2^1\|x_3\|+2^0\|x_4\|}=\no.
\end{equation}
This will become clear when we consider, for  example, rule 53184 from
Table 1. For this rule, in the third column we have
 $x_2x_2\no
\no x_1x_1x_1x_1x_2x_2\no \no \no \no \no \no $.
This means that for $x_1,x_2,x_3,x_4\in \ZZ$
\begin{align*}
\hat{f}(x_1,x_2,x_3,x_4)=\no & \quad \mbox{if} \quad \|x_1,x_2
,x_3,x_4\|=0000 \\
\hat{f}(x_1,x_2,x_3,x_4)=\no & \quad \mbox{if} \quad \|x_1,x_2
,x_3,x_4\|=0001  \\ &\cdots \\
\hat{f}(x_1,x_2,x_3,x_4)=x_2 & \quad \mbox{if} \quad \|x_1,x_2
,x_3,x_4\|=0111 \\
\hat{f}(x_1,x_2,x_3,x_4)=x_1 & \quad \mbox{if} \quad \|x_1,x_2
,x_3,x_4\|=1000 \\  &\cdots \\
\hat{f}(x_1,x_2,x_3,x_4)=x_2 & \quad \mbox{if} \quad \|x_1,x_2
,x_3,x_4\|=1111,
\end{align*}
which is a set of $16$ equations fully defining $\hat{f}$ for rule
53184. Column 3 of Table 1  represents sets of similar 16 equations in
a condensed form for all conservative 4-input rules.

In order to reduce the number of rules for further considerations, we
can exploit the fact that cellular automata rules obtained by spatial
reflection or conjugation (interchanging zeros and ones in rule
table), or both reflection and conjugation of a given rule have
similar properties as the original rule. Let, for a given $f\in
F(\{0,1\},n)$,
\begin{align}
 f_R(x_1,x_2,x_3,x_4) &=f(x_4,x_3,x_2,x_1),\\
 f_C(x_1,x_2,x_3,x_4) &=1-f(1-x_1,1-x_2,1-x_3,1-x_4),\\
 f_{RC}(x_1,x_2,x_3,x_4)&=f_C(x_4,x_3,x_2,x_1).
\end{align}
All rules in the set $\{f,f_R,f_C,f_{RC}\}$ have similar dynamics,
therefore it is enough to consider only one of them. For conservative
4-input rules, we have seven such sets, or equivalence classes:
 $\{49024,$ $59946,$ $65026,$ $43944\}$,
 $\{53184,$ $58082,$ $64524,$ $47288\}$,
 $\{56528,$ $57580,$ $62660,$ $51448\}$,
 $\{58336,$ $52930,$ $63544,$ $48268\}$,
 $\{60200,$ $48770,$ $60200,$ $48770\}$,
 $\{61680,$ $52428,$ $61680,$ $52428\}$, and
 $\{65280,$ $43690,$ $65280,$ $43690\}$.
 We will now present
mappings $g$ for the first rule in each set, defined in terms of the
step function $\Theta$ (for $m\in \ZZ$, $\Theta(m)=1$ if $m>0$,
otherwise $\Theta(m)=0$) and  the delta function (for $m,n\in
\ZZ$, $\delta_{m,n}=1$ if $m=1$, otherwise $\delta_{m,n}=0$):
\begin{align*}
 \mbox{Rule 49024:} \quad g(x_0,x_1,x_2,x_3)&=x_1-1+\Theta(x_3+x_2-2x_1-3) ,\\
 \mbox{Rule 53184:} \quad  g(x_0,x_1,x_2,x_3)&=x_1-1+\Theta(x_2-x_1-1) ,\\
 \mbox{Rule 56528:} \quad  g(x_0,x_1,x_2,x_3)&=x_1-1+\delta_{2,x_2-x_1} ,\\
 \mbox{Rule 58336:} \quad  g(x_0,x_1,x_2,x_3)&=x_1-1+\Theta(x_2-x_1-2),\\
 \mbox{Rule 60200:} \quad  g(x_0,x_1,x_2,x_3)&=x_1-1+\Theta(x_2-x_1-2)-
                          \delta_{1,x_2-x_1} \Theta(x_1-x_0-1),\\
 \mbox{Rule 61680:} \quad  g(x_0,x_1,x_2, x_3)&=x_1-1,\\
 \mbox{Rule 65280:} \quad  g(x_0,x_1,x_2,x_3)&=x_1.
\end{align*}
Note that for all the above rules, $g$ is a function of only 4
arguments (with the exception of rule 60200, it is really a function
of 3 arguments $x_1,x_2,x_3$, but for the sake of uniformity we always
used $x_0,x_1,x_2,x_3$). For all these rules, $\phi g_{\infty,1}=
\phi f_\infty$. Note that we could redefine $g$'s and introduce
$\tilde{g}(x_0,x_1,x_2,x_3)=g(x_0,x_1,x_2,x_3)+c$, where $c\in
\ZZ$, and then we would have $\phi \tilde{g}_{\infty,1}=
\phi f_{\infty,c}$.

\section{Concluding remarks}

We have demonstrated that a local mapping $f$ in  a space of
bisequences over $\{0,1\}$ which conserve the number of nonzero sites
can be viewed as a deterministic particle system evolving according to
a local mapping $g$ in a space of increasing bisequences over $\ZZ$.
We also presented an algorithm for determination of the local mapping
in the space of particle coordinates corresponding to mapping $f$.

Viewing conservative  CA as systems of interacting particles is often
very useful in solving problems related to CA dynamics, in particular
problems which can be broadly characterized as {\em forward problems}
\cite{Gutowitz91}: given a CA rule, determine (predict) its
properties. For instance, it is often easier to characterize the
trajectory $\{g_\infty^{i}(s)\}_{i=0}^{\infty}$ of a point $s \in
\tilde{X}(\ZZ)$, then the trajectory
$\{f_\infty^{i}(x)\}_{i=0}^{\infty}$ of a point $x\in X(\{0,1\}$. An
example of such approach can be found in \cite{paper11}, where we
considered properties of trajectories of configurations in simplified
deterministic models of road traffic flow.

\providecommand{\bysame}{\leavevmode\hbox to3em{\hrulefill}\thinspace}

\end{document}